# Some Aspects concerning the Cyber-Physical Systems Approach in Power Systems

N. Arghira[1], I. Fagarasan[2], G. Stamatescu[1],
S.St. Iliescu[2], I. Stamatescu[1], V. Calofir[3]

**Abstract** – The paper carries out a review of the main functional aspects of an electro-energetic system, principles which lead to an evolution or even to a paradigm change the the control of such complex systems. The repositioning of the physical and information systems, mainly through the development of the computing and communication distributed entities, lead to a new structuring of the control approach in power systems, in accordance to the modern Cyber-Physical Systems (CPS) paradigm.

## 1 Introduction

The power engineering field, in particular the electro-energetic system has been, is and will remain a key area of special technical-engineering focus for a vast array of specialists.

Defining the target approached object system, the electro-energetic system (SEE), as a collection of elements for electrical energy generation, transformation, transport and distribution, one could highlight two generic aspects which concur towards new paradigms, or eventually to the reconfiguration of existing ones, in order to design and implement control strategies for such systems:

- the characteristic of large system of the SEE;
- new technologies for electrical energy generation, in connexion with new digital technologies for measurement and control.

## 2 The Electro-energetic System – Large System

A systemic approach to SEEs highlights a series of important characteristics from the point of view of the control of such systems:

- the constituent elements of a SEE – generators, transformers, transport lines, distribution elements, form a complex, heterogenous structure, distributed over a large geographic area;
- the „internal" component elements of a SEE carry a high degree of interconnection;

---

[1] Assistant Professor, PhD at Faculty of Automatic Control and Computer Science, University "Politehnica" of Bucharest, Romania,
  e-mail:  nicoleta.arghira@aii.pub.ro,  grigore.stamatescu@upb.ro,  iulia.dumitru@aii.pub.ro
[2] Professor, PhD at Faculty of Automatic Control and Computer Science, University "Politehnica" of Bucharest, Romania,
  e-mail: ioana@ shiva.pub.ro, iliescu@shiva.pub.ro
[3] Associate Teaching Assistant, PhD at Faculty of Automatic Control and Computer Science, University "Politehnica" of Bucharest, Romania,
  e-mail: vasile@shiva.pub.ro

- the constituent elements of a SEE are defined by fast and slow processes which have to collaborate as good as possible in order to achieve the common goal for which they were designed;
- a SEE's goal is to generate electrical energy and to assure uninterrupted of consumers with this energy at the desired quality parameters (voltage and frequency);
- the vast majority of processes within a SEE are of auto-control (auto-balancing) type but having a static behaviour of unacceptable large degree, thus large stationary errors, which assumes the compulsory usage of automation equipment in order for the electrical energy delivered to the consumers to conform to the desired quality indicators;
- the increasing integration within the SEE of new renewable energy generation resources has lead to the occurence of new challenges in classic SEE's:
    - generation has become distributed, close to the consumer, where the main change in SEE control paradigm comes from;
    - the lack of controlability in energy generation, especially for wind and photovoltaic sources, large imprecision of such sources, among a context of potential climatic changes, strogly affecting the balance between generation and consumption.
- Not existing yet significant technologies for electrical energy storage, an important objective is to follow the balance between generation and consumption. One challenge has to be highlighted, which in a not-so-distant future will potentially become a new source of disturbances in a SEE. This is represented by the possibility that the number of automobiles which use electrical energy for propulsion becomes so large that at a global level we would have to account for the electrical energy stored inside the batteries of the automobiles;
- We have witnessed over the last ten years at an increase between the interconnections between the large SEE's. For example, in Europe, UTCE was built, ENTSOE at the continental level respectively. The operation within such a system poses a series of compulsory conditions which have to be complied to in terms of quantitative and qualitative indices.

## 3 Aspects concerning the interaction between new ICT technologies and the control of an SEE

The adoption of an information- and knowledge-based society concept has determined a spectacular development of digital technologies, in both the hardware and the software domains.

Following this path, new ways for implementation have emerged in the areas of control and communications.

Concepts like: *Smart Grid*, *Virtual Power Plant (VPP), new SCADA*, have appeared. To this we might add a large range of new intelligent sensors and GPS-equipped devices. Embedded networked devices are increasingly being adopted and deployed in the most prominent form of Wireless Sensor and Actuator Networks (WSAN). These are usually accompanied by the development of new robust data collection, pre-processing and communication protocols targeted at producing reliable, timely information in control applications for power systems. By exploiting the on-node local limited computation resources, the decision is brought closer to the process achieving better results in large and complex systems.

## 4 Cyber-Physical Systems and SEE

By adapting the working language to the concepts of a SEE, we will define a CPS as a highly-complex ensemble composed of a computing system – Cyber System (CS), information system, a system composed of physical elements – Physical System (PS) and a highly developed and complex communication system - Communication System (C). The proposed approach for a CPS-defined SEE is illustrated in Figure 1.

The communication system enables the coordination among CS and PS, using either wired or wireless mediums and protocols, leading to the so-called control-computing convergence.

The increase in the importance of the communication system opens up new challenges, especially concerning the vulnerability to cyber-attacks, targeted at neutralizing or worse taking control of a part or even of the whole SEE.

In principle, the challenges which can be addressed by applying the CPS concept to power systems are the following:
- cooperative real-time control of the protection operator;
- power flow control;
- qualitative and quantitative control of stability;
- the efficiency of generation and usage of electrical and thermal energy;
- control of operational safety;
- model and modeling.

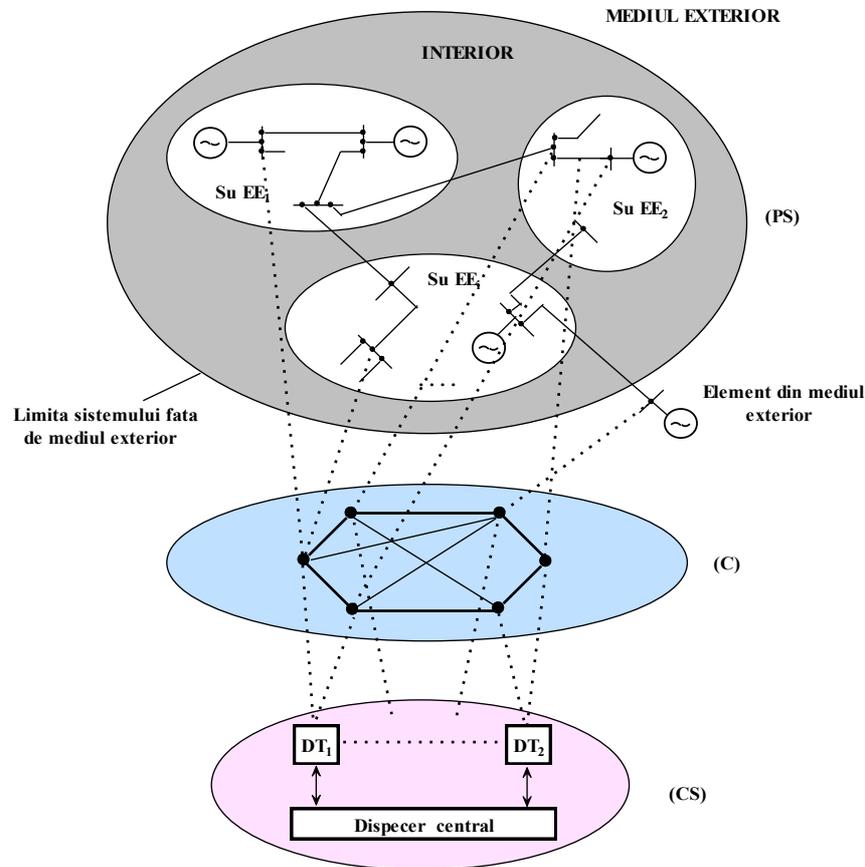

**Fig. 1.** Component subsystems of a CPS for SEE

In the CPS approach of a subsystem of the SEE, starting with the systemic concept and systems analysis, we have to highlight, using a top-down method, the three subsystems: CS, C and PS.

It has to be mentioned that according to the above-mentioned concepts, a preliminary identification of the object system has to be carried out.

Figure 2 shows an example of this concept, where as subsystem of a SEE, the automatic control system for frequency and active power was considered (SRAf-P).

The decomposition among the three subsystems leads to specific control approaches for each of the subsystems. For the PS, the problem is to determine the analytical model, which can be solved through an experimental identification procedure using step test signals. For the C subsystem a data transmission system was implemented using error-detecting codes, i.e. parity control. With regard to the CS, the control law was implemented in software with adaptive possibilities for the controller gain.

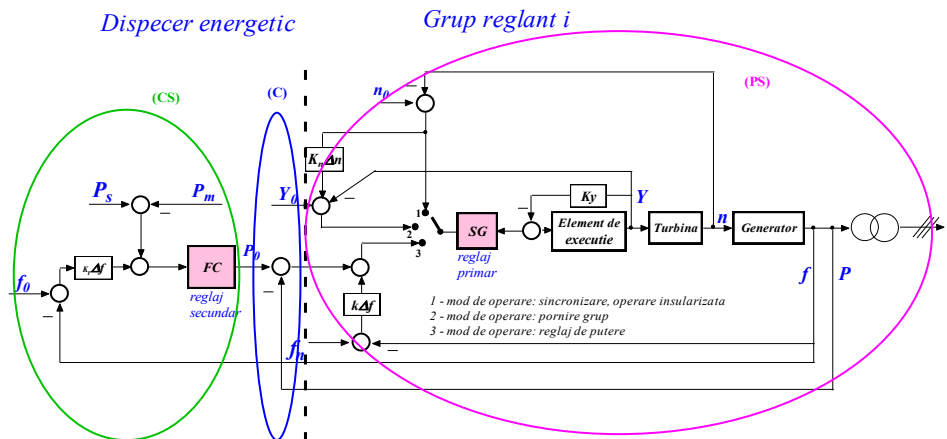

**Fig. 2.** CPS for power systems implementation

## References

1. Iliescu S. St.: Control of Power Systems. Concepts and Implementation, Tutorial Session, AQTR – THETA 16, Cluj-Napoca, Romania (2008).

2. Morris T., Sristava A., Reaves B., Pavurapu K., Abdelwahed S., Vaughn R., McGrew W., Dandass Y.: Engineering Future Cyber-Physical Energy Systems: Challenges, Research Needs, and Roadmap, North American Power Symposium, NAPS, pp. 1-6, Starkville, USA (2009).

3. Stamatescu I., Stamatescu G., Arghira N., Fagarasan I., Iliescu S. St.: Fuzzy Decision Support System for Solar Tracking Optimization, Proceedings of the 12th International Conference on Development and Application Systems (DAS), Suceava, Romania (2014).

4. Fagarasan I., Iliescu S.St., Arghira N., Asan V.: Integration Issues for Wind Generation Units, Proceedings of the 1st Workshop on Energy, Transport and Environment Control Applications, pp. 53-61, Bucharest, Romania (2010).

5. Costianu D.R., Iliescu S.St., Arghira N., Fagarasan I., Dumitru I.: Distributed generation influence on Smart Distribution Grids, ICSP – IFAC Proceedings, Vol. 2, pp. 54-57, Cluj-Napoca, Romania (2013).

6. Arghira N., Costianu D.R.: Virtual Power Plant Operation and Control, UNITE Doctoral Symposium, pp. 118-121, Bucharest, Romania (2011).

7. Stamatescu G., Stamatescu I., Arghira N., Fagarasan I., Iliescu S. St.: Embedded Networked Monitoring and Control for Renewable Energy Storage Systems, IEEE DAS, pp. 1-6, Suceava, Romania (2014).
5